\documentclass{article}

\usepackage[final]{nips_2016}

\usepackage[utf8]{inputenc} 
\usepackage[T1]{fontenc}    
\usepackage{url}            
\usepackage{booktabs}       
\usepackage{amsfonts}       
\usepackage{nicefrac}       
\usepackage{microtype}      
\usepackage{graphicx}

\title{Survival Prediction with Limited Features: a Top Performing Approach from the DREAM ALS Stratification Prize4Life Challenge}

\author{
  Christoph Kurz \thanks{twitter: @pihive}\\
  Helmholtz Zentrum M\"unchen\\
  Institute of Health Economics and Health Care Management\\
  Neuherberg, Germany \\
  \texttt{christoph.kurz@helmholtz-muenchen.de} \\
}

\begin{document}

\maketitle

\begin{abstract}
Survival prediction with small sets of features is a highly relevant topic for decision-making in clinical practice.
I describe a method for predicting survival of amyotrophic lateral sclerosis (ALS) patients that was developed as a submission to the DREAM ALS Stratification Prize4Life Challenge held in summer 2015 
to find the most accurate prediction of ALS progression and survival. ALS is a neurodegenerative disease with very heterogeneous survival times. Based on patient data from two national registries, solvers were asked to predict survival for three different time intervals, which was then evaluated on undisclosed information from additional data.

I describe methods used to generate new features from existing ones from longitudinal data, selecting the most predictive features, and developing the best survival model. 
I show that easily obtainable engineered features can significantly improve prediction and could be incorporated into clinical practice.
Furthermore, my prediction model confirms previous reports suggesting that past disease progression measured by the ALSFRS (ALS functional rating scale score), time since disease onset, onset site, and age are strong predictors for survival.
Regarding prediction accuracy, this approach ranked second. 
\end{abstract}

\section{Introduction}
Amyotrophic Lateral Sclerosis (ALS), also known as Lou Gehrig's syndrome or Motor Neuron Disease, is an incurable degenerative disease of the motor nervous system: it involves progressive and irreversible damage or degeneration of nerve cells (neurons) responsible for muscle movements. Symptoms include speech and swallowing disorders, impaired coordination, and weakness of the arm and hand muscles leading to progressive reduction in the activities of daily living and eventually death. 
The site of disease onset as experienced by the patient can be a limb ("limb onset") or the muscles controlling speaking and swallowing ("bulbar onset"), or occasionally both.
ALS is a rare disease with a prevalence of 3-8 per 100,000. The median survival in ALS from first symptom ranges from 2 to 4 years~(\cite{czaplinski2006predictability}), but a proportion of patients survive more than 10 years (Prof. Stephen Hawking being the most prominent example). This heterogeneity of ALS outcomes makes survival difficult to predict~(\cite{louwerse1997amyotrophic, gordon2013predicting}).

To improve prediction a new computational community challenge has been presented within the Dialogue on Reverse Engineering Assessment and Methods (DREAM) project~(\cite{stolovitzky2007dialogue}). 
The DREAM ALS Stratification Prize4Life Challenge 2015 (hereafter referred to as "the challenge") is a follow-up to the DREAM Phil Bowen ALS Prediction Prize4Life Challenge~(\cite{kuffner2015crowdsourced}) with the aim of finding predictors of disease progression that can be used to aid clinical care, identify new disease predictors, and potentially significantly reduce the costs of future ALS clinical trials. 

The challenge provided two different data sets from different trials and registries, and participants were asked to predict survival, given as probability of death, for three time points. A restricting part of the challenge was to use only six variables from the data set provided.

In this paper, I present my approach to feature engineering, feature selection, and the survival model.

\section{Materials and Methods}
\subsection*{Data description}
Data sources are the PROACT (Pooled Resource Open-Access ALS Clinical Trials) clinical trial database~(\cite{proact}) and a mix of two national registries (NatReg). All data were collected between 1990 and 2015. The PROACT database provides a training set of $N=7200$ patients, combined from an unknown number of publicly and privately conducted ALS clinical trials; the registries contain $N=986$ patients from two national registries, Italy and Ireland. Algorithms will be evaluated on data ($N=1900$) from six new trials not yet available in PROACT, which may contain variables dissimilar to the training set. Therefore, an additional set of $N=200$ records from this validation set was provided. Additionally, a set of $N=493$ patients from the NatReg set was withheld for validation. Both PROACT and NatReg data sets provide a variety of baseline and longitudinal values, such as laboratory measures, medication, and vital status. Overall, more than 100 variables were available. According to the organizers, this is the largest data set ever used for the prediction of ALS progression and survival.

\subsection*{Data cleaning and feature engineering}
Based on the longitudinal data from the 0--3 months period, I generated new variables for each of the original longitudinal variables. Table~\ref{table1} presents an overview. Note that some of these new variables are highly correlated, e.g., if there is only one observation for a patient, then \emph{mean}, \emph{max}, \emph{min}, \emph{first}, and \emph{last} will be the same value, while \emph{sd}, \emph{diff}, \emph{minmax} and \emph{slope} will be set to zero. I therefore chose to delete variables with a Pearson correlation coeffiecient of $\rho>0.95$. 

\begin{table}[!ht]
	\caption{
		{\bf Description of all additional variables that were generated from each longitudinal variable.}}
	\begin{tabular}{|l|l|}
		\hline
		\multicolumn{1}{|l|}{\bf Variable suffix} & \multicolumn{1}{|l|}{\bf Description}\\ \hline
		mean & The mean of all values in the 0--3 months period \\ \hline
		sd & The standard deviation of all values in the 0-3 months period \\ \hline
		max & The maximum value in the 0--3 months period \\ \hline
		min & The minimum value in the 0--3 months period \\ \hline
		diff & The difference between the maximum and the minimum values\\ \hline
		first & The first observed value in the 0--3 months period\\ \hline
		last & The last observed value in the 0--3 months period \\ \hline
		len & The number of observed values in the 0--3 months period \\ \hline
		minmax & The slope of the max and min regression line \\ \hline
		slope & The slope of the regression line in the 0--3 months period \\ \hline
	\end{tabular}
	\label{table1}
\end{table}

\subsection*{Feature selection}
Even after removing those variables with $\rho>0.95$, some of the variables are still highly correlated; for example, the \texttt{hand} and \texttt{leg} items from the ALSFRS score are strongly related. Therefore, traditional stepwise elimination approaches are not efficient~(\cite{hastie2005elements}). I based variable selection on three different algorithms: random forest variable importance, gradient boosting, and elastic net regularization. Both boosting and elastic net regularization are built upon a Cox proportional hazards (PH) model. 

\subsection*{Survival model}
To find the best survival model for my data, I compared five different algorithms in a 5-fold cross-validation setting:
\begin{enumerate}
	\item Recusive partitioning for survival trees (RPART), as described in the book by Breiman~(\cite{breiman1984classification}) and implemented in the \texttt{rpart} package.
	\item The random survival forest (RSF) approach illustrated by Ishwaran et al~(\cite{ishwaran2008random}) and implemented in the accompanying R package \texttt{randomForestSRC}. 
	\item The Cox PH model, implemented in the \texttt{survival} package~(\cite{survival-package}).
	\item A Cox PH model with Elastic Net regularization, implemented in the \texttt{glmnet} package~(\cite{glmnet1, glmnet2}).
	\item A gradient boosted Cox model, implemented in the \texttt{CoxBoost} package.
\end{enumerate}

\section{Results and Discussion}

\begin{figure}[!ht]
	\centering
	\includegraphics[width=0.99\textwidth]{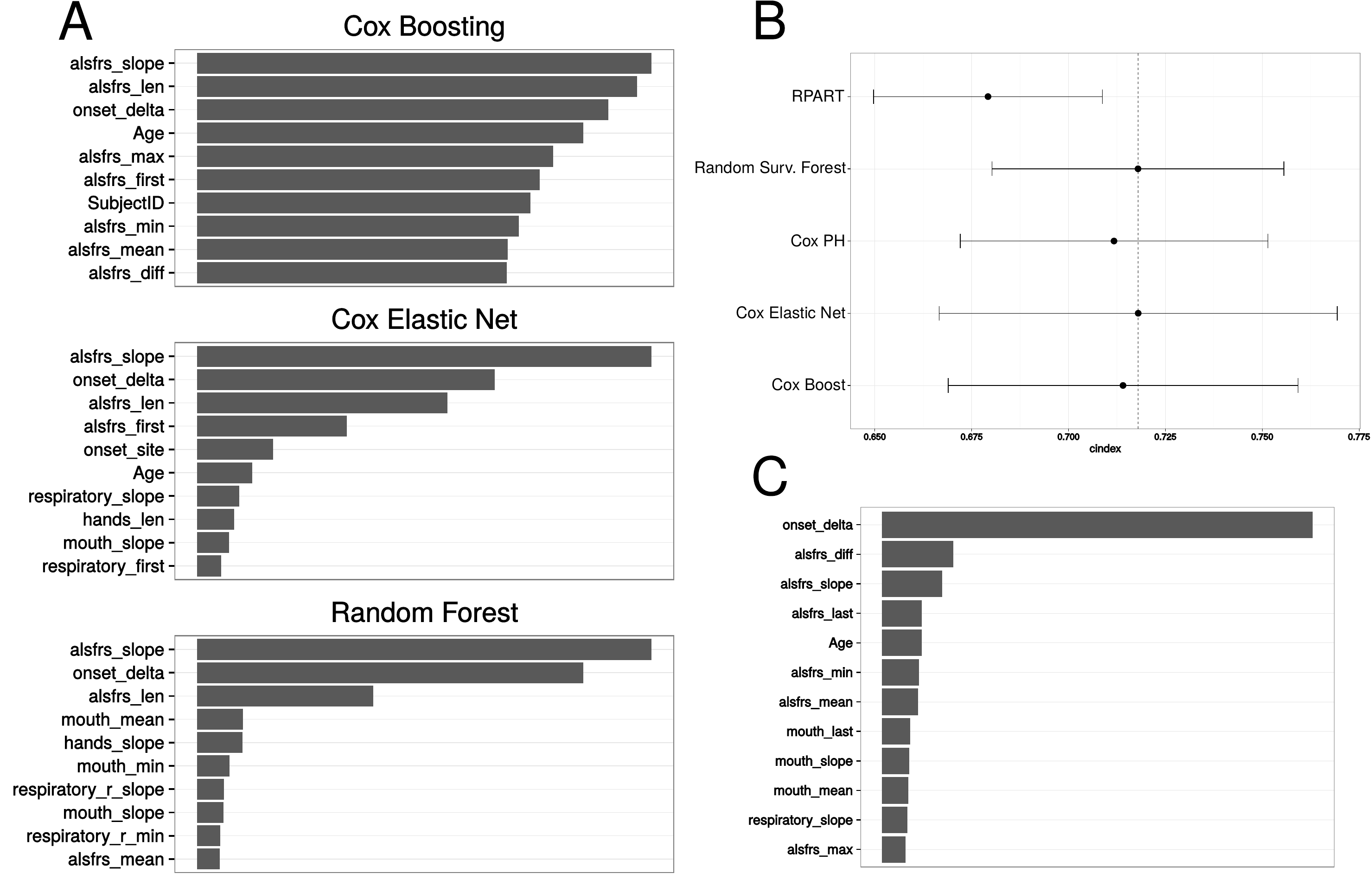}
	\caption{\textbf{(A)} Variable importance plots for the 10 most important variables in all three models. \textbf{(B)} Cross-validation results of the survival model comparison. Bars represent the lower and upper limits of c-index in the folds; black dot is the mean c-index across all folds. \textbf{(C)} Variable importance plots for the final model.}
	\label{fig:plot2}
\end{figure}

\subsection*{Selected variables}
Fig~\ref{fig:plot2} (A) shows the variable importance plots of all three models.
Variables generated by the \texttt{alsfrs} feature have the greatest importance in all three methods. \texttt{respiratory} (the score for the respiratory item in the ALSFRS) variables are regarded as of almost equally high importance, whereas \texttt{onset\_delta} (the time since disease onset in days) was only chosen by random forest and boosting. The other variables I included in the model were \texttt{mouth} (the score for the mouth item in the ALSFRS), \texttt{Age} (baseline age of each patient), and \texttt{onset\_site} (site of disease onset). According to the literature~(\cite{kollewe2008alsfrs}), the forced vital capacity is a predictive feature for survival. However, it could not be used in the model because of its large number of missing values.
Interestingly, the team that was placed first in the challenge selected exactly the same variables.

\subsection*{Survival model}
For the final model, I chose RSF as it showed the best performance in the cross-validation. One advantage of RSF is that it does not
expect linear features or even features that interact linearly. It is robust to outliers and
can handle missing values. Furthermore, RSF works very well with large numbers of observations and has a natural way of dealing with data of mixed type (e.g., categorial
variables, count variables, or binary variables).
Fig~\ref{fig:plot2} (B) shows the cross-validation model comparison.

\subsection*{Variable importance}
The 12 most important variables in the final model are depicted in Fig~\ref{fig:plot2} (C). Except for the baseline variables \texttt{onset\_delta} and \texttt{Age}, all other variables are engineered. 
The time of disease onset (\texttt{onset\_delta}) is by far the most important variable because early onset corresponds with slow progression and therefore increased survival.
The difference between the maximum and minimum measured ALSFRS score in the 0--3 months period (\texttt{alsfrs\_diff}) is considered to be the second most important variable. This feature covers those patients with rapid degeneration in the first 3 months, but it cannot detect patients who recover in this period as it does not include the time points of the measurement. Interestingly, it is still slightly favored over \texttt{alsfrs\_slope}, which measures the slope of the ALSFRS linear regression line and can thus cover patients whose health improved. I speculate that the combination of both leads to improved predictions, and this is well handled in the arms of the decision trees. For example, positive slope and large difference mean rapid recovery, whereas negative slope and large difference result in rapid degeneration. Other highly important features such as the last (\texttt{alsfrs\_last}) and the smallest (\texttt{alsfrs\_min}) measured ALSFRS score make sense intuitively. The site of disease onset (\texttt{onset\_site}) is not among the most important predictors, but again may play a role in combination with other variables. 

\section{Conclusion}
In this article, I have described the approach that was placed second in the DREAM ALS Stratification challenge. I have shown how feature engineering can be used when limited data are available in longitudinal form. New features generated from existing variables can be more predictive than others already at hand. The generated features can easily be incorporated into a clinical setting. I have also shown that variable selection should be examined by a variety of methods, as each method can produce different results. I have explained my choice of the final survival model.

\small

\bibliographystyle{plainnat}
\bibliography{dream_als_nips}

\end{document}